\pdfoutput=1
\documentclass[12pt]{iopart}
\usepackage{graphicx,epstopdf}
\usepackage{caption}
\usepackage{subcaption}
\usepackage{gensymb}
\usepackage{textcomp}
\usepackage{xcolor}
\usepackage[binary-units=true]{siunitx}

\begin{document}

\title{An EDUGATE simulation toolkit based on the educational easyPET}

\author{P.M.M Correia$^1$, J. Menoita$^1$, A.L.M Silva$^1$, N. Romanyshyn$^1$, and J.F.C. A Veloso$^1$}
\address{$^1$ I3N - Department of Physics, University of Aveiro, Campus Universit\'{a}rio de Santiago, 3810-193, Aveiro, Portugal}

\ead{pmcorreia@ua.pt}
\vspace{10pt}

\begin{abstract}
EasyPET is a new concept of a Positron Emission Tomography (PET) scanner using an innovative acquisition method based on two rotation axes for the movement of detector pairs. Due to its simplicity, it is suitable for education purposes, to teach students about the PET technology and its basic concepts, from the radiation detecting and analogue pulse analysis to the coincidence sorting and image reconstruction. The concept allows achieving high and uniform position resolution over the whole field of view (FoV), by eliminating parallax errors due to the depth of interaction (DoI), which are typical of ring-based PET systems, so quality images are obtained even without state-of-the-art image reconstruction algorithms. The technology developed at the University of Aveiro with a patent-pending, is licensed to CAEN S.p.A, and included in the educational catalogue of the company. In this work, a simulation toolkit based in the Edugate platform was developed to simulate the EasyPET system. It can simulate all the physical aspects of the product, such us the scanning range, variable Field-of-View (FOV), scintillator energy resolution, coincidence time and energy window, among others. A simple image reconstruction algorithm based on Filtered-back-projection (FBP) is implemented. The toolkit allows a quick analysis in classroom of the simulation results. The platform was also used to study the new EasyPET 3D version, and a simulation of a NEMA NU 4-2008 IQ phantom was performed, demonstrating the capability of the platform not only for education purposes but also for research.

\textbf{Patent} Universidade de Aveiro: PCT/IB2016/051487
\end{abstract}




\maketitle

\section{Introduction}
Positron Emission Tomography (PET) is a powerful imaging technique in nuclear medicine, widely used for clinical diagnostics, for example in early stage cancer detection, but also in preclinical studies, for development and testing of new diagnostic and therapeutic agents, prior to clinical trials \cite{Yao01092012}.
Convectional PET scanners have ring geometries, where the scintillation detectors are distributed uniformly in a full (or partial) circunference. The stack of several circumferences allows 3D imaging. More exotic geometries, like parallel plates or boxes, are also used for both preclinical and clinical applications\cite{0031-9155-62-15-6207,0031-9155-42-12-012}.

\begin{figure}[hbtp]
\centering
\includegraphics[scale=0.7]{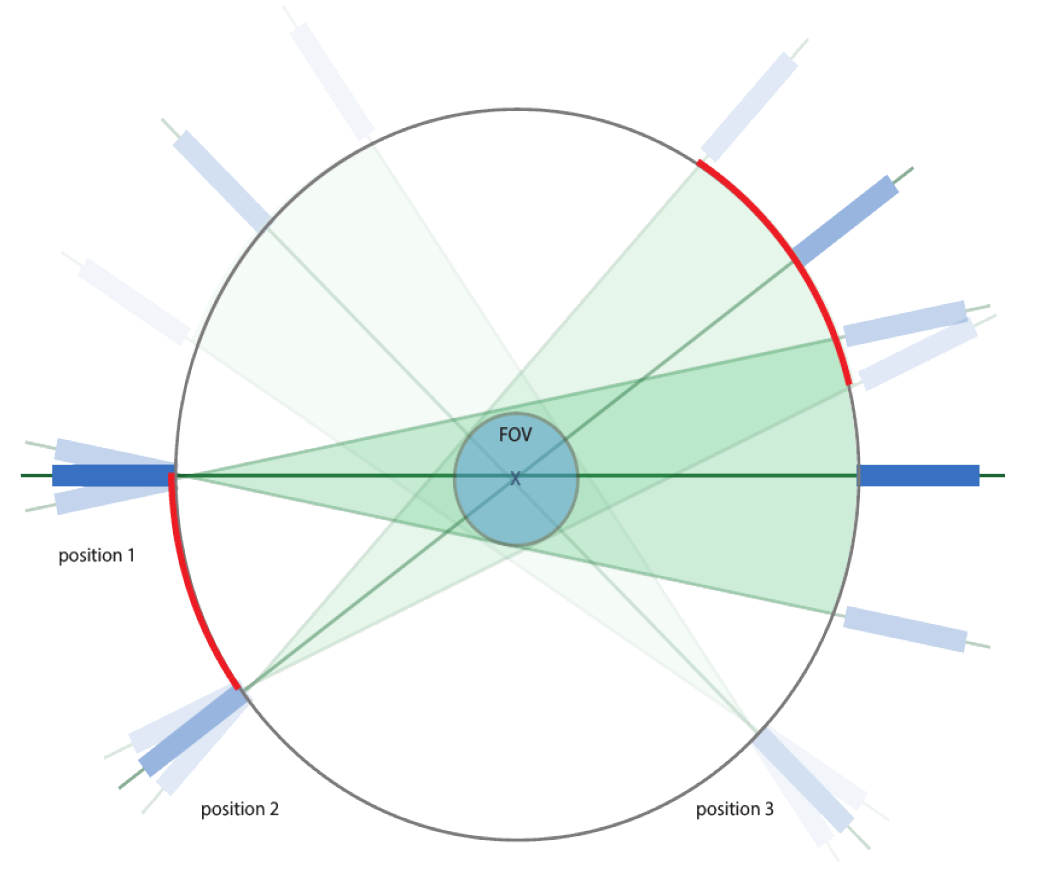}
\caption{EasyPET working principle scheme. Three positions are represented, in each one the scanner acquires a fanbeam projection. The FOV dimension is defined by the angular aperture of the fanbeam scan.\label{fig:easyPET_method}}
\end{figure}

\begin{figure}
    \centering
    \begin{subfigure}[c]{0.4\textwidth}
        \includegraphics[width=\textwidth]{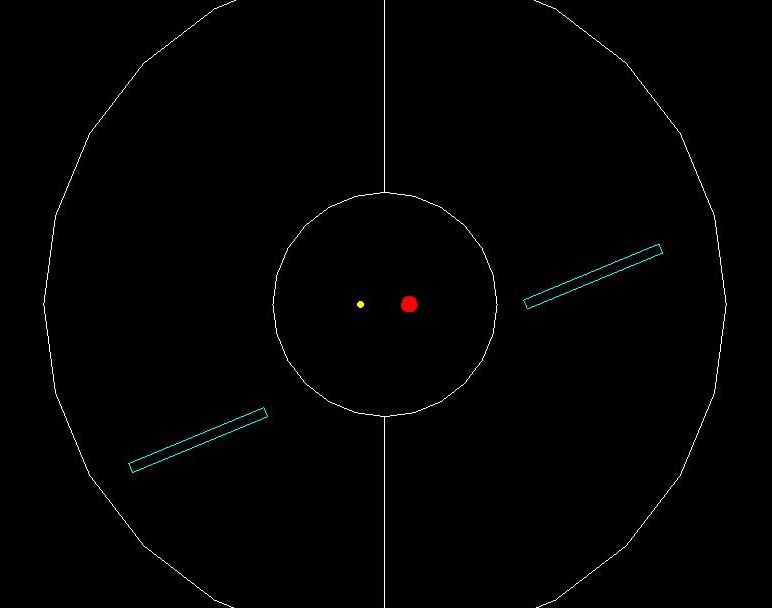}
        \caption{}
        \label{fig:GateImage2D}
    \end{subfigure}
    ~
    \begin{subfigure}[c]{0.3\textwidth}
        \includegraphics[width=\textwidth]{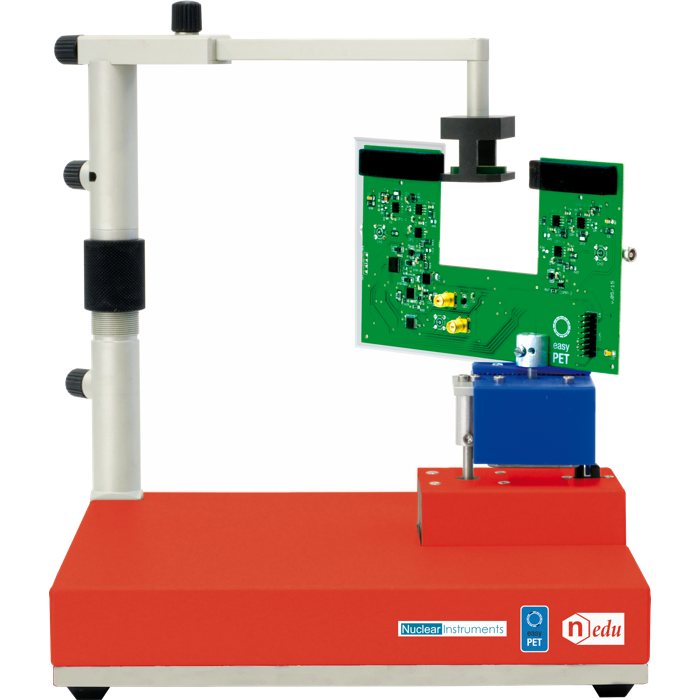}
        \caption{}
        \label{fig:Fig2b_easyPET_big}
    \end{subfigure}
    \caption{(a) Top view of a GATE simulation of EasyPET. Two radioactive sources (different sizes) are placed in the FOV. (b) Picture of the EasyPET system available from CAEN S.p.A. }\label{fig:EasyPET_pic}
   
\end{figure}

Included in this exotic geometries of PET scanners is the EasyPET system developed at the University of Aveiro. This PET system is based in an innovative method in which two detecting cells rotate according two axes, performing a full
2D scan \cite{AROSIO2017644,AROSIO2016} similar to the one obtained with a ring based PET scanner.A scheme of the EasyPET working system is depicted in \fref{fig:easyPET_method} and \fref{fig:EasyPET_pic}. 
The acquistion method was patented by the University of Aveiro \cite{calapez2016positron} and the technology was licensed to CAEN S.p.A., being now a commercial product, part of the CAEN educational catalog \cite{caenspa2016}.

The EasyPET system
mimics a fanbeam acquisition method present in many CT
scanners, but instead of use a full array of scintillation detectors, only
two are used and a fan rotation is used to cover the entire FOV. Each time a coincidence is detected between the two scintillation detectors, the line-of-response (LOR) considered for the two annihilation photons trajectory is the line between the two front faces of the scintillation detectors. In this configurations, the parallax effect is avoided\cite{AROSIO2016,AROSIO2017644}. The reason is that photons always penetrate the scintillation detectors perpendicularly to the front face, avoiding the uncertainty in the LOR determination.  The parallax effect is always a concern in traditional PET scanner configurations, due to the uncertainty in the Depth-of-Interaction (DOI) of the gamma photons that enter the scintillation detectors with an oblique angle. Although several solutions to minimize the effect have been applied successfully\cite{0031-9155-53-17-R01,0031-9155-61-18-6635,0031-9155-60-9-3673,pcorreia2014}, it has big influence on the final image if no correction is applied.

To simulate the behaviour of scanners used in Nuclear Medicine, a new software named GATE (Geant4 Application for Tomography Emission)\cite{1236960} has been developed by the OpenGate collaboration. The software was built on top of Geant4\cite{AGOSTINELLI2003250} engine, already widely validated in High-Energy physics. 
GATE is now a very mature software, that was used for simulations on different techniques besides PET, such as Single Photon Emission Computed Tomography (SPECT), Computed Tomography (CT), Optical Imaging, among others.  

As part of the GATE platform, an educational section is available, called EduGATE\cite{PIETRZYK201365}. It contains practical miscellaneous examples of how to use GATE for educative purpose.

In this work, a new EduGATE example is presented, dedicated to the EasyPET system. It is capable of keep up with the
real EasyPET product in terms of image acquisitions, allowing to predict its behaviour and understand the basis of a simplified PET scanner.
For these reasons, it is an interesting complement for those who already have access to EasyPET but also for those who don't but want to use the simulation capabilities to study or teach PET technology.

\subsection{EasyPET Working principle}

\begin{figure}
    \centering
    \begin{subfigure}[b]{0.3\textwidth}
        \includegraphics[width=\textwidth]{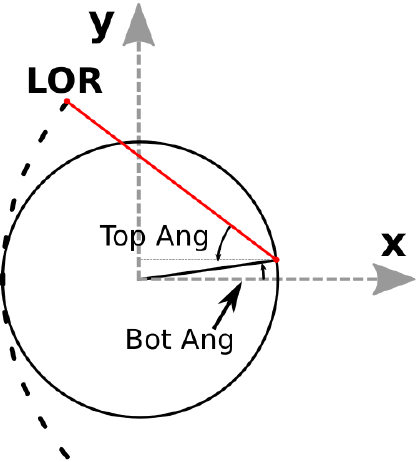}
        \caption{}
        \label{fig:rotating_angles}
    \end{subfigure}
    ~
    \begin{subfigure}[b]{0.3\textwidth}
        \includegraphics[width=\textwidth]{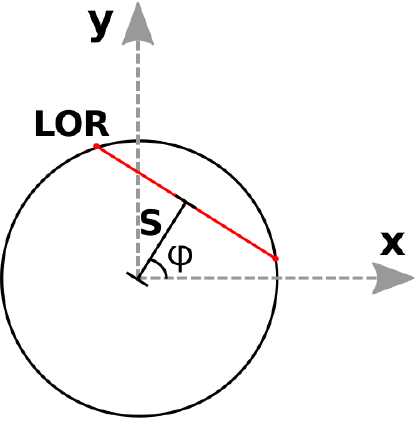}
        \caption{}
        \label{fig:angle2sino}
    \end{subfigure}
    \caption{(a) Two rotating angles ($Top Angle$ and $Bottom Angle$) are needed to perform a valid acquisition. (b) Angular conversion from rotating angles to sinogram coordinates $S$ and $\phi$.}\label{fig:angles}
\end{figure}

The innovation about the EasyPET acquisition system is related with the number and movement of the scintillation detectors. Instead of use a full ring array of scintillation detectors, only two cells are used, separated by a distance $d$. To cover all FOV, two rotation movements, applied at different axis of rotation, are needed, as depicted in \fref{fig:angles}:
\begin{itemize}
\item \textbf{Fan rotation} - rotation axis is located on the front face of one the scintillation detectors. The angular arc covered by this rotation is defined the \textbf{Top Range} angle, while the angular increments are defined by the \textbf{Top Angle}. 
\item \textbf{Axial rotation} - rotation axis is located on the center of the FOV. For each full fan rotation, an increment in the axial rotation is performed, by an angle called \textbf{Bottom Angle}
\end{itemize}   
       
In order to perform image reconstruction, a sinogram can be constructed from the angles Bottom Angle and Top Angle using \eref{eq1} and \eref{eq2}:
\begin{eqnarray}
\phi &= 90\degree + Bottom Angle +  Top Angle \label{eq1}\\
S & = sin(Top Angle) \times d/2 \label{eq2} 
\end{eqnarray}
where $S$ is the distance of the LOR to the center of the projection axis, $\phi$ is the angle between the projection axis and the x-axis and $d$ is the distance between the scintillation detectors.

\section{Methods}{\label{methods}}

\subsection{Simulation Toolkit}
\begin{figure}
    \centering
    \begin{subfigure}[c]{\textwidth}
    \centering
        \includegraphics[height=.45\textwidth]{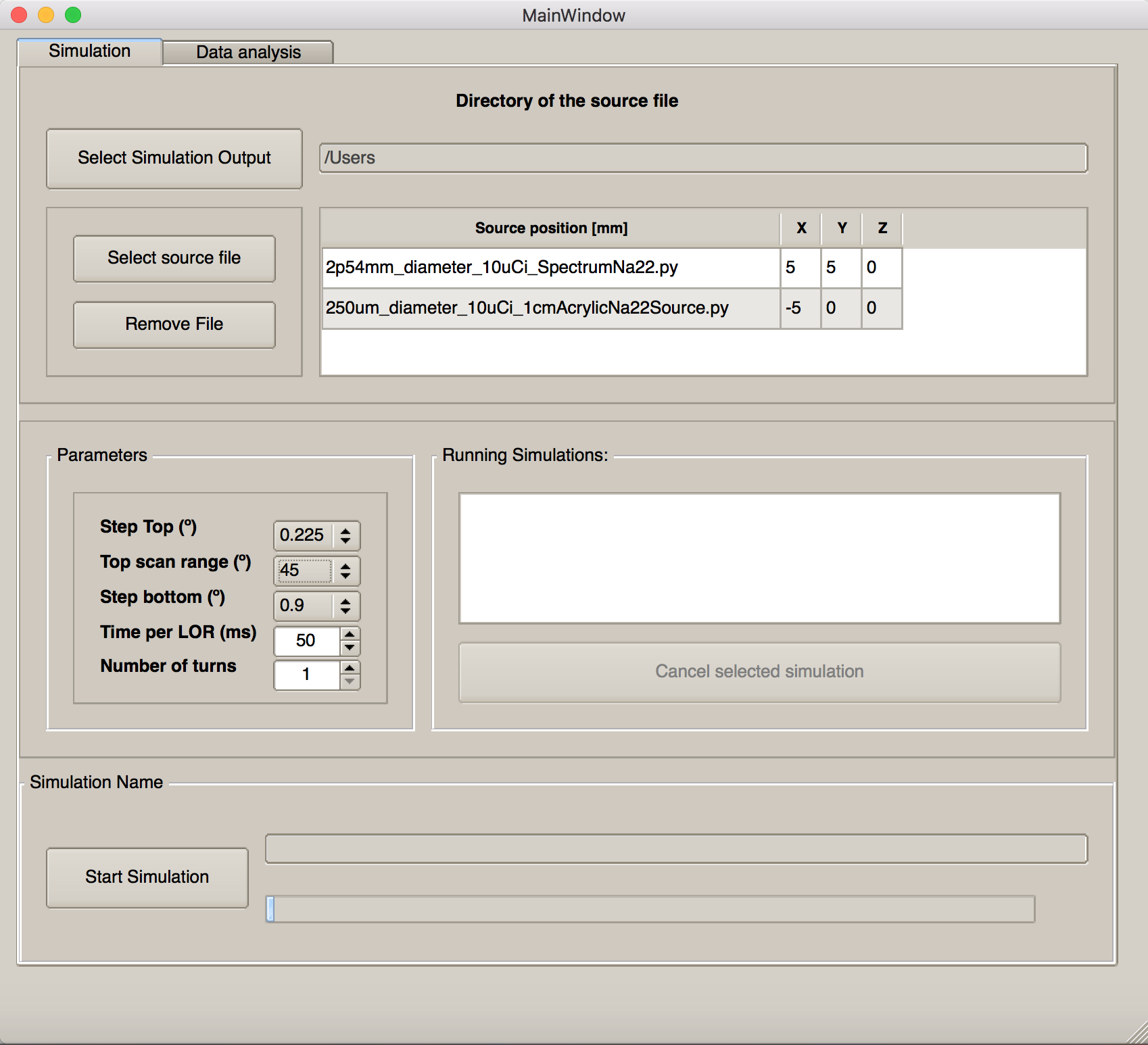}
        \caption{}
        \label{fig:rotating_angles2}
    \end{subfigure}
    \begin{subfigure}[c]{\textwidth}
    \centering
    \includegraphics[height=.45\textwidth]{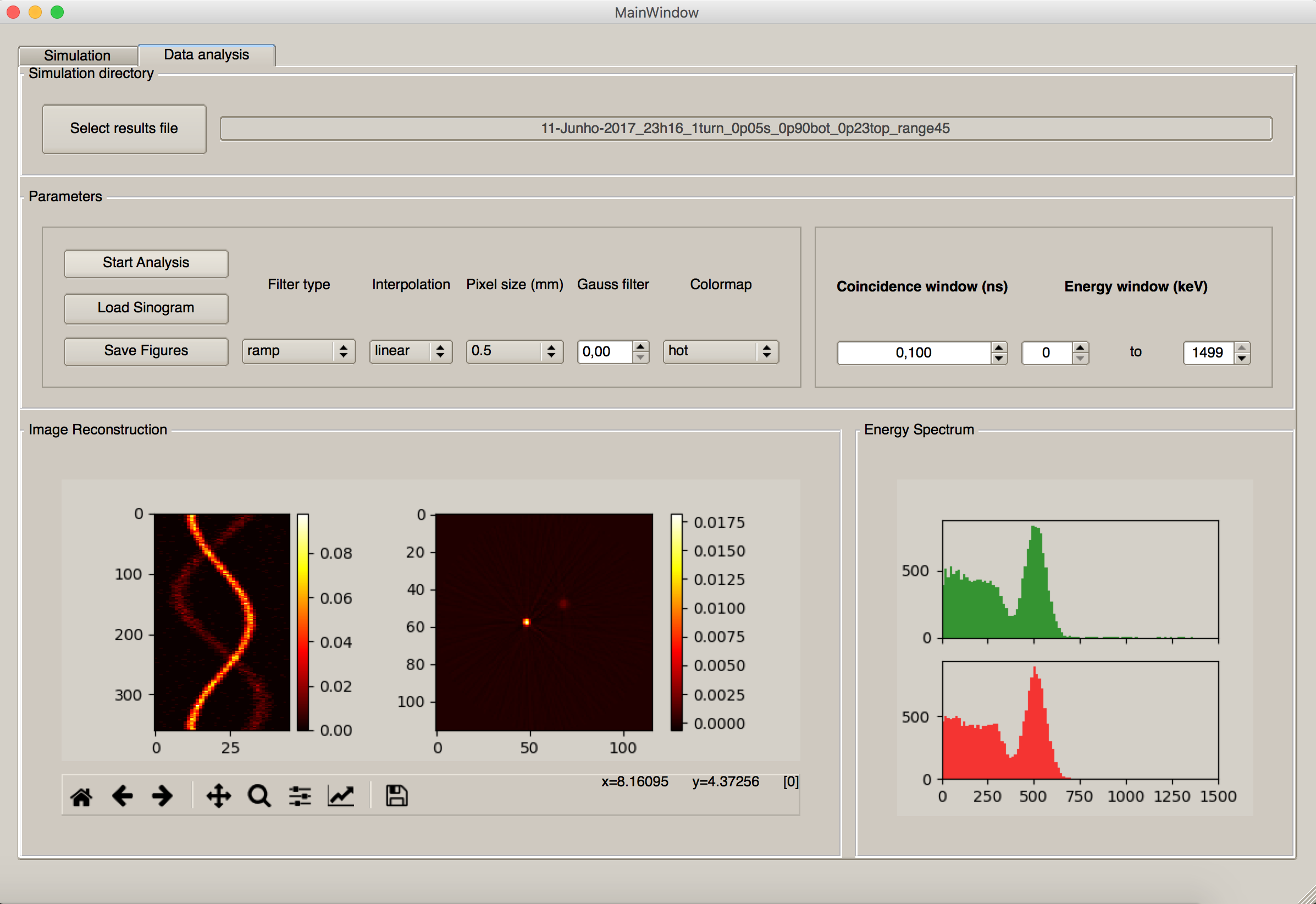}
        \caption{}
        \label{fig:angle2sino}
    \end{subfigure}
    \caption{GUI developed in Qt4 interfaced with Python using PyQt.  Two $^{22}$Na Spectrum sources (a disk with 2.54 mm of diameter and 10 $\mu$Ci and a sphere with 0.25 mm of diameter and 10 $\mu$Ci), at different locations, are shown as an example. Two menus are available:\textbf{ a)} one to define the simulation parameters and \textbf{b)} the other to reconstruct the resulting images\label{fig:pythonUI}}
    \end{figure}

A simple Graphical User Interface (GUI), depicted in \fref{fig:pythonUI}, was developed using Qt4\footnote[7]{https://www.qt.io}, and interfaced with Python using PyQt\footnote[6]{https://riverbankcomputing.com/software/pyqt/intro}
Two menus are available in the GUI. \textit{Simulation} menu handles the generation of the .mac files that will be used for the GATE simulation. The second menu, \textit{Data Analysis}, is meant to allow the user to analyse the output data and reconstruct the images.

In the \textit{Simulation} menu, the user has several parameters that can control, in order to mimic an experimental acquisition. 
The first parameter that need to be selected is the simulation output path where the simulation files will be available. After that, the radioactive source must be chosen. Different pre-defined sources are available (a $^{22}$Na point-like sphere  and a 2.54 mm disk, a $^{18}$F NEMA NU 2008 IQ phantom, etc). The number of sources their correspondent position in the simulation volume are also user-defined.
Before generate the simulation files, the movement parameters related with the acquisition itself must be selected, as described below:
\begin{itemize}
\item
\textbf{Step Top} is the minimum angular separation between the positions of the detecting cell in the fan arc during the fan rotation movement. By default this is selected as 0.9\degree.

\item \textbf{Top scan range} defines the number of arc positions that will be executed. It must be a multiple of the Step Top defined before. 

\item
\textbf{Step Bottom} is the angular separation between two consecutive axial positions. The number (integer value) of fan views to be executed during the acquisition is given by \textbf{360\degree /Step Bottom}. 

\item 
\textbf{Time per LOR} is the amount of time, in miliseconds, that the mechanic system will count coincidences in each LOR. 

\item \textbf{Number of Turns} allows the repetition of more than one turn around the axial axis. It is used for long acquisitions in order to study dynamic processes, where fast turns are needed to acquire images at different moments.

\end{itemize}

When clicking in \textbf{Start Simulation} button, simulation files are created and simulation starts running using a pre-installed version of GATE, accessible in the computer. Simulations running in a given moment are displayed in the \textbf{Running Simulation} list, and are removed from once they finish. 

\subsection{Simulation output}
Simulation results are stored in a Root\cite{BRUN199781} file, located in the directory path selected in the \textit{Simulation} menu. 
 The raw data is stored in a form of two Root trees, named \textit{SinglesScanner1} and \textit{SinglesScanner2}, each one corresponding to one detecting cell. The interaction time, position and energy of each gamma photon detected, among other variables, are recorded. Each entry of the tree corresponds to an event where a photon interacted with a detecting cell. 

In the \textit{Data Analysis} menu, \textit{Start Analysis} button will run in background a Root script that converts the raw data of the simulation to a sinogram.   

To extract the coincidences from the raw data to a sinogram, a coincidence sorter was developed. 
For each event in one of the detecting cells, the coincidence sorter searches event in the other cell that occurred within the previously defined coincidence window. Once a coincidence is found, the corresponding sinogram coordinates are calculated and the sinogram updated with a new coincidence.  
The file containing the sinogram is then stored with the same name as the .root file but with extension .sinogram.
The data in this file is organized as a matrix containing $N\times M$ integers, where $N$ is the number of $\phi$ positions, $M$ the number of $S$ positions and each integer number is the number of coincidenes detected for each pair ($S$,$\phi$).

\subsection{Image reconstruction method}
A Filtered-Backprojection (FBP) algorithm distributed by the Python library scikit-image\cite{scikit-image} was implemented for fast image reconstruction.
The use of external software for image reconstruction is then avoided, although data can be exported for external image reconstruction software if the user needs.
Some parameters for the  reconstruction can be selected by the user:
\begin{itemize}
\item \textbf{Filter type} to be used in the FBP reconstruction can be selected among Ramp, Shepp-Logan, Cosine, Hamming, Hann or no filter. 
\item \textbf{Interpolation} mechanism determines the type of interpolation applied to the sinogram by the FBP algorithm.
\item \textbf{Pixel size} defines the output size of the image. A small pixel size allows the visualization of smaller details, at the expenses of increase significantly the noise of the image and also the computing time.
\item \textbf{Gaussian filter} - when the pixel size starts to be very small, the use of a Gaussian filter to the sinogram usually smooths the image and remove some of the noise. However, large values for this filter usually hides small details in the image.
\item \textbf{Colormap} change is possible directly from the GUI, as well as the \textbf{Colorbar limits} (using both left and right button of the mouse directly in the colobar). This option can enhancing details that would be otherwise hidden.
\end{itemize}

Changing any of this parameters will automatically update the image and display the new results. There is also the possibility to save the images (sinogram and reconstruction) in image format, for a latter analysis, if desired.  

\section{Results and discussion}{\label{results_and_discussion}}
\subsection{Position Resolution}
To validate the simulation toolkit, different acquisitions were performed. Results are presented in this section. 

\begin{figure}[hbtp]
\centering
\includegraphics[width=0.75\textwidth]{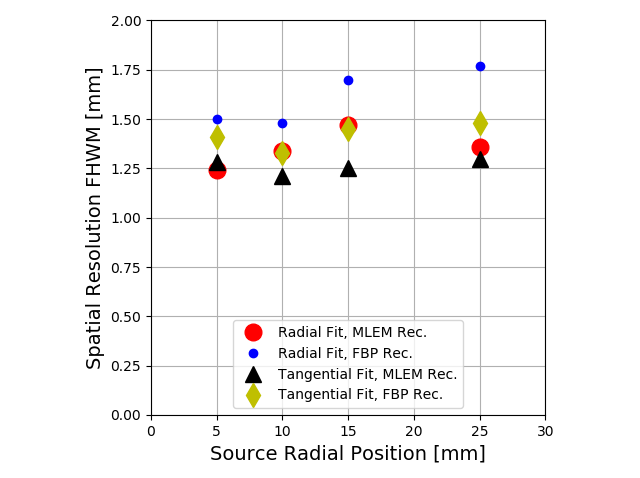}
\caption{Spatial resolution for a point-like source located along the radial position of the scanner. Results using FBP and MLEM\cite{pedrosa} reconstruction methods are compared. No visible degradation is observed, due to the non presence of parallax effect.\label{fig:spatial_resolution}}
\end{figure}

To evaluate the position resolution of the easyPET system, following NU 4-2008 standards methodology for measuring scanner performance parameters for small-animal PET\cite{NEMA2008}, a point-like $^{22}$Na source embedded in a 1 cm$^3$ cube of PMMA was used. The position resolution was evaluated for various positions along the radial direction. For each acquisition, the number of recorded coincidences was approximately 5000, the images where reconstructed using FBP. The spatial resolution is measured as the FWHM obtained from the gaussian fit to the line profile crossing the center of the source, in the radial and tangential directions. The results are depicted in \fref{fig:spatial_resolution}. 

Simulation results for spatial resolution measurement are in good agreement with experimental measurements using easyPET, and are also well compared with other PET scanners using scintillators with 2x2 sections. 

\subsection{Coincidence time and energy window}

The coincidence sorter developed allows to define a coincidence time window from 1 ps up 100 $\mu$s. This allows to observe the effect of large coincidence time windows for the number of random coincidences and the consequent image quality degradation. 

\begin{figure}[ht]
\centering
\includegraphics[width=0.75\textwidth]{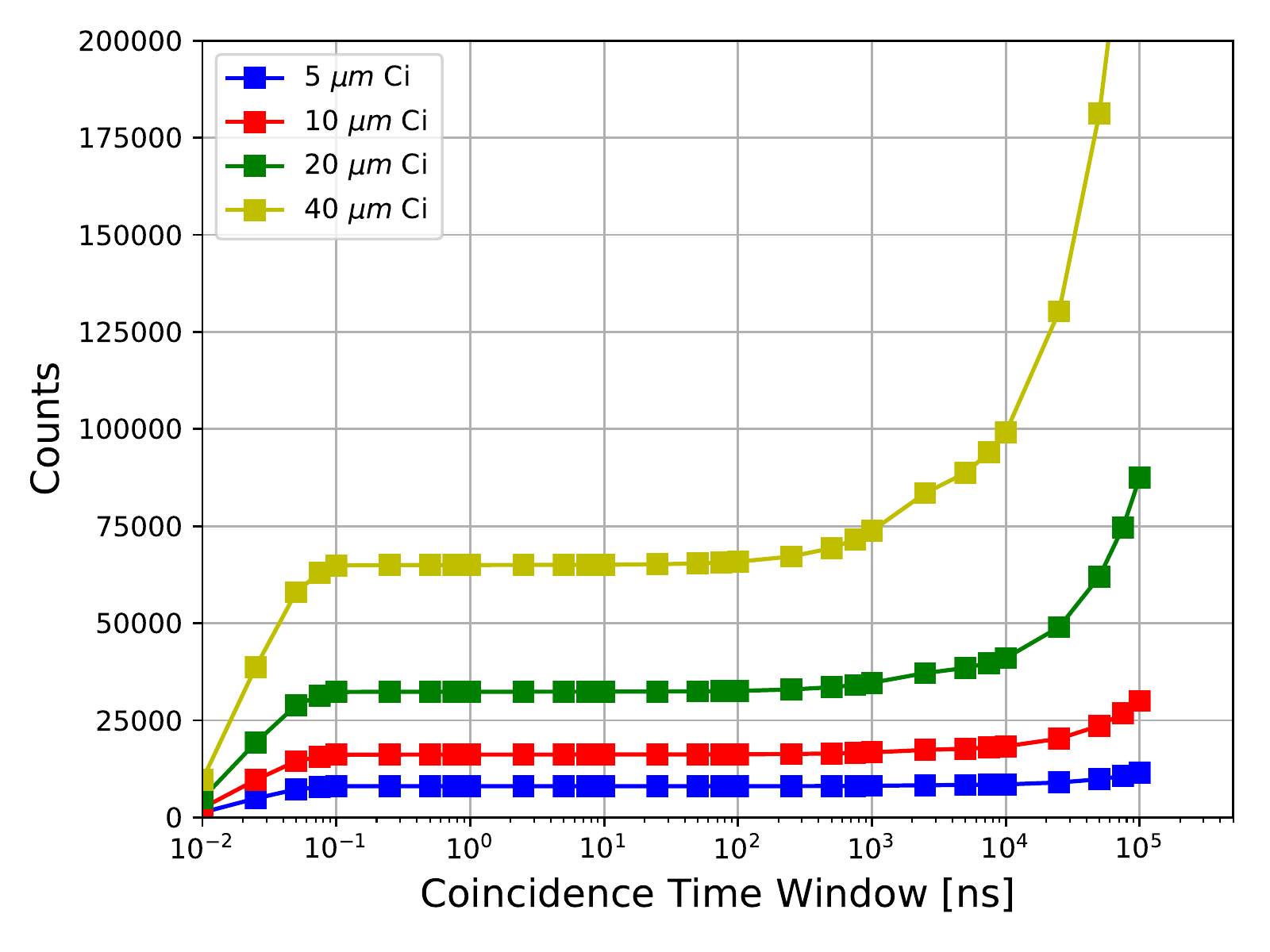}
\caption{Number of coincidences detected for different source activities as a function of the coincidence time window. A well defined plateau is appears, when the coincidence time window is enough for detect almost all the true coincidences and discards non-true concidences. \label{fig:counts_vs_coincidence_window}}
\end{figure}

To evaluate the number of coincidences detected as a function of the coincidence time window, an acquisition was simulated with a $^{22}$Na sphere of 0.25 mm diameter embedded in a 1 cm$^3$ cube of PMMA, with variable activity (from 5 to 40 $\mu$Ci), and in the center of the FOV, using \textbf{Top Angle} = 0.225$\degree$,  \textbf{Top Range} = 45$\degree$, \textbf{Bottom Angle} = 0.9$\degree$, \textbf{Timer per LOR} = 50 ms and 1 turn. Results are depicted in \fref{fig:counts_vs_coincidence_window}.
When the time window rises up to 100 ps, the number of detected coincidences first increases exponentially and then reaches a plateau. The first increase is explained by the number of true coincidences that were being discarded due to the extremely short time windows (below 100 ps).
The distance travelled by one photon in 100 ps is about 3 cm, which is the length of the crystal simulated. Due to that, a photon interaction near the front face of the crystal is detected up to 100 ps before a similar interaction near the rear face of the crystal (not considering the delays due to scintillation and the optical photons travelling inside the crystal), explaining why true coincidences are lost when small time windows are considered in the simulations. 

For values above 100 ns, for the case with higher activity, random coincidences start to appear, having a negative effect in the image quality. This increase is observed for higher coincidence time windows when lower activities are used. 

For this simulations we considered sources with relatively low activity (up to 40 $\mu$Ci) and only two crystals. Although the left part of the plateau region is reached for the same time windows, independent of the activity and number of crystals, the second increase of the number of considered coincidences, here observed for time windows above 100 ns, is dependent of the events rate (higher activity means higher probability to detect random coincidences)

    \begin{figure}[ht]
    \centering
        \includegraphics[width=\textwidth]{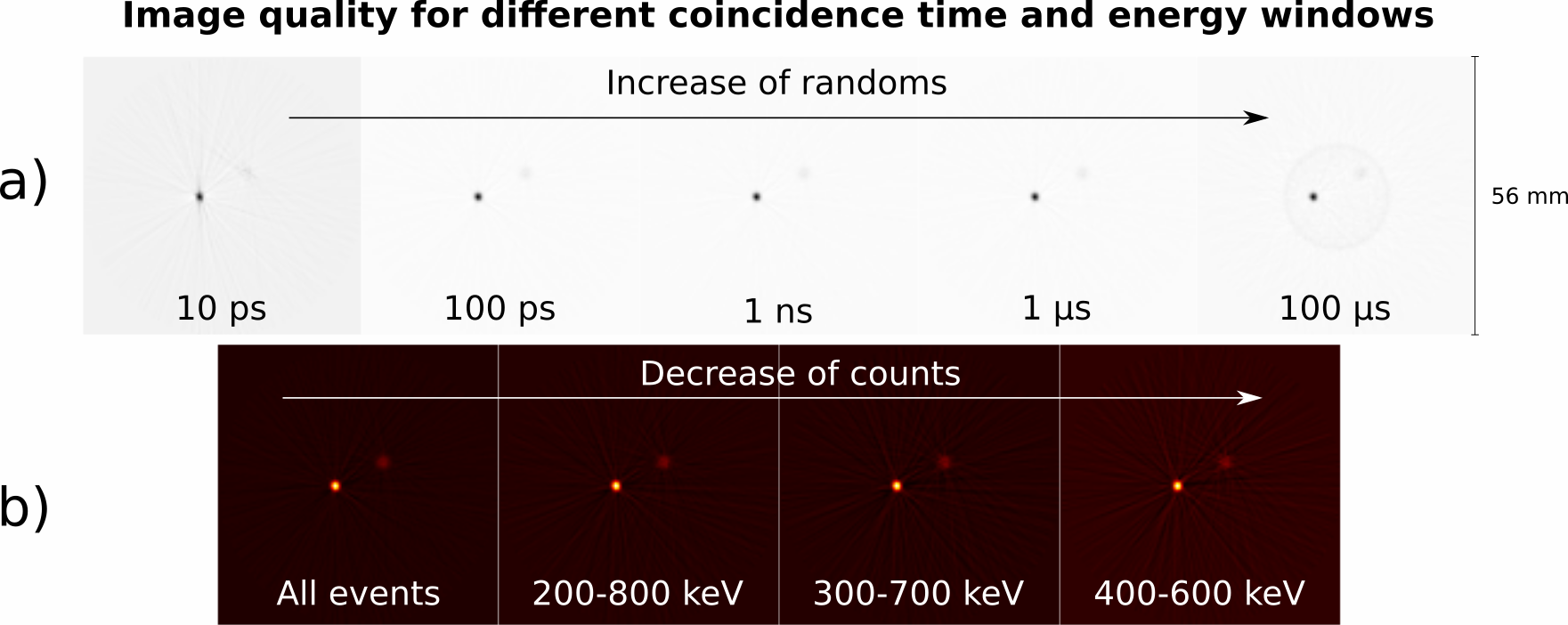}
 \caption{Effect of \textbf{a)} the coincidence time and \textbf{b)} energy windows for the final image. For very narrow time windows the resulting image is poor, having low statistic, and for large windows, the noise due to randoms deteriorates the results. Also in this case the energy selection of \SI{511}{\kilo\electronvolt} events lowers the statistics.\label{fig:plot_vs_coincidence_window}}
    \end{figure}

The plot depicted in \fref{fig:plot_vs_coincidence_window} shows the reconstructed images of a different simulation, with two $^{22}$Na sources (a disk with 2.54 mm
of diameter and 10 uCi and a sphere with 0.25 mm of diameter and 10 uCi), at diferent
locations ((5,5,0) mm and (-5 0 0) mm, respectively), using \textbf{Top Angle} = 0.225$\degree$,  \textbf{Top Range} = 45$\degree$, \textbf{Bottom Angle} = 0.9$\degree$, \textbf{Timer per LOR} = 50 ms and 1 turn, for different coincidence time windows and energy cuts. For very short time windows (below 100 ps), the loss of counts is the main factor for the image quality degradation, while no visible degradation is seen for large coincidence windows until this window remains below a certain value (activity dependent).
For higher coincidence time windows, the random coincidences introduce a noticeable noise in the final image. 

In terms of energy windows selection, since in this situation we are simulating
low activity sources (bellow 100 $\mu$Ci) the positive effect of the increase of the ratio
between true and random coincidences that is achieved through the energy selection
of only the \SI{511}{\kilo\electronvolt} events is not noticeable. In fact, the image quality decreases when
an extreme energy cut (\num{400}-\SI{600}{\kilo\electronvolt}) is applied to the events, due to the reduction
of the number of coincidences and consequently decrease of the S/N ratio (\fref{fig:plot_vs_coincidence_window}b)).

\subsection{Image Reconstruction}

\begin{figure}[ht]
    \centering
    \begin{subfigure}[c]{0.45\textwidth}
        \includegraphics[width=\textwidth]{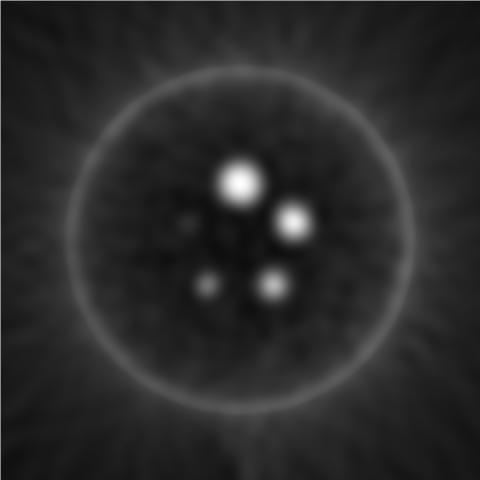}
        \caption{}
        \label{fig:Fig8_capillars}
    \end{subfigure}
    \begin{subfigure}[c]{0.45\textwidth}
        \includegraphics[width=\textwidth]{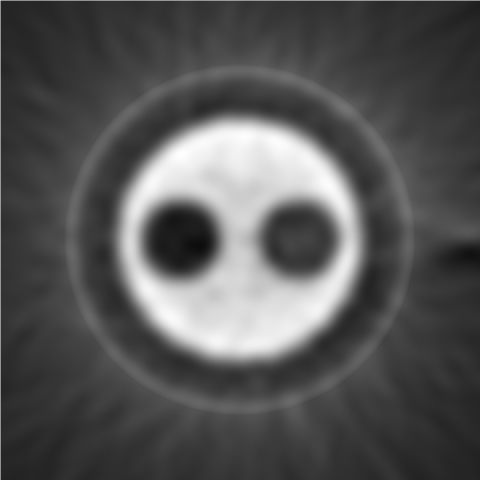}
        \caption{}
        \label{fig:Fig8_vessels}
    \end{subfigure}
    \caption{Simulated images of the NEMA NU 4-2008 IQ phantom using the Edugate platform adapted for the EasyPET 3D version. Images reconstructed using FBP algorithm.\label{fig:nema2008_simulation}}
    \end{figure}

The EasyPET edugate toolkit, initially developed to simulate the educational version of the EasyPET, was adapted for the 3D version of the device, which is currently under characterization, and used to simulate the NEMA NU 4-2008 IQ phantom\cite{NEMA2008}. Results shown in \fref{fig:nema2008_simulation} demonstrate the potentiality of the platform to simulate real PET scanners and to be used as a powerful educational toolkit, combined with the other examples available in the Edugate toolkit.

\section{Summary and Conclusion}{\label{summary_and conclusion}} 

The easyPET, developed at the University of Aveiro, is a simplified and affordable PET scanner, developed for students, available for purchase from the CAEN educational kit. The main goal of the system is to allow students to perform and understand simple experiments in PET, from the electric signals circuitry to the final image reconstruction of radioactive sources. 

A simulation toolkit based on the EasyPET system is presented in this work. It is an important tool to understand the EasyPET operation principles, and all the geometric parameters related with data acquisition. Using the toolkit, students can change the acquisition parameters, such as the angular steps, the FOV, the speed of the acquisition, the coincidence time window and the energy threshold for the detected $\gamma$ photons.  
The type, number and position of the radioactive sources are also selected by the user.

Student  can test a FBP algorithm for image reconstruction with multiple input parameters, and evaluate their impact in the final reconstructed image. 

Three experiments are suggested in this paper: the evaluation of the position resolution of the system using a point-like $^{22}$Na source, the effect of the coincidence time window and energy window for the reconstructed image.

The toolkit has demonstrated very similar results comparing with the physical EasyPET system. Spatial resolution bellow \SI{1.5}{\milli\meter} was obtained, which is a good result compared with PET scanners with the same crystal size.

The toolkit was adapted for the upgraded 3D version of the EasyPET, and the results of an simulation using the NEMA IQ phantom, demonstrates the capabilities and flexibility of the platform not only for educational purposes, but also for research. 
Soon available as an EduGate module, this toolkit will be useful not only for students that already have the EasyPET system available at their college, which can compare simulated results with the experimental acquisitions, but also for other students that do not have access to the EasyPET but can benefit in classes.

\ack
This work was partially supported by project POCI-01-0145-FEDER- 016855 and PTDC/BBB-IMG/4909/2014, and project easyPET CENTRO-01- 0247-FEDER-017823, CENTRO2020, COMPETE, FEDER, POCI and FCT (Lisbon) programs.
P.M.M. Correia was supported by FCT (Lisbon) grant
PD/BD/52330/2013 and by I3N laboratory, funded by
UID/CTM/50025/2013. 

\section*{References}
\bibliographystyle{iopart-num}
\bibliography{article_bibtex}

\providecommand{\newblock}{}
\begin{thebibliography}{10}
\expandafter\ifx\csname url\endcsname\relax
  \def\url#1{{\tt #1}}\fi
\expandafter\ifx\csname urlprefix\endcsname\relax\def\urlprefix{URL }\fi
\providecommand{\eprint}[2][]{\url{#2}}

\bibitem{Yao01092012}
Yao R, Lecomte R and Crawford E~S 2012 {\em Journal of Nuclear Medicine
  Technology\/} {\bf 40} 157--165

\bibitem{0031-9155-62-15-6207}
Kyme A~Z, Judenhofer M~S, Gong K, Bec J, Selfridge A, Du J, Qi J, Cherry S~R
  and Meikle S~R {\em Physics in Medicine and Biology\/} {\bf 62} 6207

\bibitem{0031-9155-42-12-012}
Freifelder R and Karp J~S {\em Physics in Medicine and Biology\/} {\bf 42} 2463

\bibitem{AROSIO2017644}
Arosio V, Caccia M, Castro I, {Correia} P~M~M, Mattone C, Moutinho L, Santoro
  R, Silva A and Veloso J 2017 {\em Nuclear Instruments and Methods in Physics
  Research Section A: Accelerators, Spectrometers, Detectors and Associated
  Equipment\/} {\bf 845} 644 -- 647

\bibitem{AROSIO2016}
Arosio V, Caccia M, Castro I, Correia P, Mattone C, Moutinho L, Santoro R,
  Silva A and Veloso J 2016 {\em NSS-MIC 2016 Conference Record\/}

\bibitem{calapez2016positron}
Veloso J~F~C~A, Castro I~F~C, Moutinho L~M~C, Carramate L~F~N~D, Correia P~M~M
  and Silva A~L~M 2016 Patent - "positron emission tomography system and method
  with two rotation shafts" ref. {PCT} /IB2016/051,487
  \urlprefix\url{http://www.google.com/patents/WO2016147130A1?cl=en}

\bibitem{caenspa2016}
CAEN 2016 {SP5700} easypet
  http://www.caen.it/csite/CaenProd.jsp?idmod=1025\&parent=61 last visited at
  29-07-2017

\bibitem{0031-9155-53-17-R01}
Lewellen T~K {\em Physics in Medicine and Biology\/} {\bf 53} R287

\bibitem{0031-9155-61-18-6635}
Uchida H, Sakai T, Yamauchi H, Hakamata K, Shimizu K and Yamashita T {\em
  Physics in Medicine and Biology\/} {\bf 61} 6635

\bibitem{0031-9155-60-9-3673}
Cabello J, Etxebeste A, Llosá G and Ziegler S~I {\em Physics in Medicine and
  Biology\/} {\bf 60} 3673

\bibitem{pcorreia2014}
{Correia} P~M~M, {Castro} I~F~C and {Veloso} J~F~C~A 2014 {\em Second
  International Conference on Applications of Optics and Photonics\/} {\bf
  9286}

\bibitem{1236960}
Santin G, Strul D, Lazaro D, Simon L, Krieguer M, Martins M~V, Breton V and
  Morel C 2003 {\em IEEE Transactions on Nuclear Science\/} {\bf 50} 1516--1521
  ISSN 0018-9499

\bibitem{AGOSTINELLI2003250}
Agostinelli S {\em et~al.\/} 2003 {\em Nuclear Instruments and Methods in
  Physics Research Section A: Accelerators, Spectrometers, Detectors and
  Associated Equipment\/} {\bf 506} 250 -- 303

\bibitem{PIETRZYK201365}
Pietrzyk U, Zakhnini A, Axer M, Sauerzapf S, Benoit D and Gaens M 2013 {\em
  Zeitschrift für Medizinische Physik\/} {\bf 23} 65 -- 70 ISSN 0939-3889

\bibitem{BRUN199781}
Brun R and Rademakers F 1997 {\em Nuclear Instruments and Methods in Physics
  Research Section A: Accelerators, Spectrometers, Detectors and Associated
  Equipment\/} {\bf 389} 81 -- 86 ISSN 0168-9002 new Computing Techniques in
  Physics Research V

\bibitem{scikit-image}
van~der Walt S, {S}ch{\"o}nberger J~L, {Nunez-Iglesias} J, {B}oulogne F,
  {W}arner J~D, {Y}ager N, {G}ouillart E, {Y}u T and the scikit-image
  contributors 2014 {\em PeerJ\/} {\bf 2} e453 ISSN 2167-8359

\bibitem{NEMA2008}
 2018 Nu 4-2008: Performance measurements of small animal positron emission
  tomographs \textit{NEMA Standard Publication}

\bibitem{pedrosa}
S{´a} P~M~M~S 2017 Image reconstruction algorithm implementation for the
  easypet: a didactic and pre-clinical pet system
  \url{http://hdl.handle.net/10451/31700}

\end{thebibliography}

\end{document}